# DFT-based Conformational Analysis of a Phospholipid Molecule (DMPC)

S. Krishnamurty*, M. Stefanov[a], T. Mineva, S. Begu, J.M. Devoisselle, A. Goursot
Institut Gerhardt, UMR 5253 CNRS, Ecole de Chimie de Montpellier, 34296 Montpellier Cédex 5, France.
R. Zhu, D.R. Salahub
Department of Chemistry and Institute for Biocomplexity and Informatics, University of Calgary, 2500 University Drive NW, Calgary, Alberta, Canada T2N 1N4

Abstract

The conformational space of the dimyristoyl phosphatidylcholine (DMPC) molecule has been studied using Density Functional Theory (DFT), augmented with a damped empirical dispersion energy term (DFT-D). Fourteen ground-state isomers have been found with total energies within less than 1 kcal/mol. Despite differences in combinations of their torsion angles, all these conformers share a common geometric profile, which includes a balance of attractive, repulsive and constraint forces between and within specific groups of atoms. The definition of this profile fits with most of the structural characteristics deduced from measured NMR properties of DMPC solutions. The calculated vibrational spectrum of the molecule is in good agreement with experimental data obtained for DMPC bilayers. These results support the idea that DMPC molecules preserve their individual molecular structures in the various assemblies.

## I- Introduction

Phospholipids are the building blocks of biological membranes and for this reason their structures and properties have been studied extensively. Experimental studies have recognized that the knowledge of the structure and dynamics of a phospholipid monomer within an assembly is essential for understanding the bilayer functional role in bio-membranes [1-6]. Based on X-Ray structures of several types of crystallized phospholipids [7], various NMR studies [2-6] have aimed at elucidating the molecular structure of phospholipids in fluid bilayers (liquid-crystalline phase). Despite these experimental efforts, there is not yet a clear consensus about their individual structural and dynamical properties in the gel and in the less ordered liquid-crystalline states. The effects of the intra- and inter-molecular electronic forces on the structure and energetics of these systems make the task non trivial both experimentally and computationally.

Restricting the scope of this paper to dimyristoyl phosphatidylcholine (DMPC), which is one of the most studied phospholipids, we briefly summarize below the experimental information about its conformation and dynamics in different phases. In the crystallized phase, two DMPC molecules in the asymmetric unit cell of the monoclinic crystal ($P2_1$ space group) were identified, each molecule being associated with two $H_2O$ molecules at the polar head side [7]. These two conformers show different headgroup orientations associated with different torsion angles. Their coexistence in the crystal was attributed to two different but energetically similar states of the freely rotating headgroup. Therefore, taking into consideration earlier NMR results for the gel and liquid-crystalline phases of lecithin [8], similarities in the monomer head and glycerol backbone were recognized for the solid, gel and liquid-crystalline phases [7]. This conclusion is not confirmed by more recent NMR studies addressing the question of the degree of similarity between the conformations in the liquid-crystalline and single crystal structures [3-6]. The presence of two rotational isomers (rotamers A and B) obtained by rotation around the C2 - C3 bond (Figure 1) while keeping a parallel alignment of the alkyl chains, was proposed based on $^{1}H$ spin coupling values in solution at different concentrations, i.e. at monomeric and micellar states [3]. The two DMPC conformers present in the crystal unit cell correspond both to the same rotamer structure (rotamer A in ref.3). The interchange between rotamers was further assumed, but, due to the relatively long time scale of the NMR spectroscopy, this proposal cannot be verified [3].

More recent two-dimensional NMR measurements of hydrated DMPC indicate strong dipolar couplings of the β-chain carboxyl $^{13}C$ (C21 in Figure 1) with headgroup protons as well as with the phosphate $^{31}P$, suggesting thus a rigid backbone and a headgroup bent toward the beginning of the β-chain [4]. This conclusion of a rigid glycerol backbone is in contradiction with the previous proposal of rotamers interchanging in the fluid bilayer. These NMR results, taken together with other NMR data reporting significant $^{31}P$ - $^{13}C$ coupling [1-3], exclude the presence of a DMPC conformer with a head pointing away from the alkyl chains as it is the case for one of the single crystal conformers. The fact that the DMPC conformers in the single crystal structure differ from those present in the liquid-crystalline phase was also confirmed by $^{13}C$ and $^{31}P$ CP-MAS [5] and by $^{31}P$ and $^{2}H$ NMR conformational studies [6].

The data just reviewed show that the intrinsic rigidity (or flexibility) of the glycerol backbone part of the DMPC monomer is controversial, but that there is a consensus concerning structural differences between the fluid-like and the crystal conformers. This fact questions the relevance of using the single crystal atomic positions for fluid phase simulations wherein the internal monomer properties appear to be better preserved, due to the relatively small intermolecular interactions [5,6]. This may not be valid in the crystalline state due to the presence of stronger electrostatic intermolecular interactions [5].

Fourier transform infrared (FTIR) spectroscopy has also been used extensively for studying DMPC bilayers, in their crystalline, gel and liquid-crystalline lamellar phases. The use of isotopically labelled atoms has led to the assignment of vibrations of the functional groups. The presence of water has been related to red shifted vibrations of the phosphate and glycerol groups, whereas blue shifted stretching $CH_2$ vibrations have been found for temperatures above the phase transition temperature and associated with the appearance of gauche conformations in the alkyl chains [9-13] Despite the large amount of information concerning the structural organization of the lamellar crystalline, gel or liquid phases, the conformational structure of PC lipids at the molecular level has not been elucidated.

Understanding the properties and functions of membrane constituents is also a challenge for molecular simulations. Due to the large size of phospholipid assemblies, quantum mechanical (QM) modeling has been so far limited to exploring head group properties : (i) methylphosphate ions as models with semi-empirical or Hartree-Fock methods [14-16]; (ii) the solvent effect on phosphoethanolamine and phosphocholine groups with the Hartree-Fock method [17]; (iii) lipid head hydration of methylphosphocholine with Density Functional Theory (DFT) [18,19]. Very recently, the interaction of the dipalmitoyl phosphatidyl choline molecule with dipyridamole has been studied, using the semi-empirical PM3 method [20].

In contrast, classical molecular dynamics (MD) methods have been used extensively to study phospholipid bilayers. Force field parameters have been developed for lipids with various parametrizations, using QM calculations on small model molecules and reproducing experimental properties [21-25]. Among the large number of MD studies devoted to phospholipid structural properties, dihedral angle values [20,21], head group flexibility [26,27], tail orientations [24], phase changes [28,29], hydration effects [30,31], interaction with ions [25] and molecules [32,33], evaluation of local order parameters [34,35] have been explored.

Although all these theoretical studies, in conjunction with experimental techniques, provide insight into some aspects of the conformational and dynamical behaviour of phospholipids in gel and liquid-crystalline phases, a complete and coherent understanding of the electronic factors contributing to the molecular conformation and energetics is still lacking. In the present work, we thus propose to quantify, based on a QM method, the relationships between the conformational structures and the intramolecular interactions of a DMPC molecule. The analysis of the calculated vibrational spectrum shows that the main structural characteristics derived in this study allow one to reproduce the experimental Infra-red (IR) spectra. The calculated normal modes are in a very good agreement with the available experimental assignments.

## II- Methodology

The DMPC molecule has been studied with Density Functional Theory (DFT), using a linear combination of atomic orbitals, as implemented in the deMon2k program [36]. All calculations were performed using the revised PBE exchange functional (revPBE) [37] and the LYP [38] correlation functional, augmented by a damped empirical correction for dispersion-like interactions [39], referred to as DFT-D. This DFT-D approach is necessary to account for the stabilizing interaction between the alkyl chains [40].

DFT-optimized double zeta plus valence polarization (DZVP) basis sets [41] were employed for all atoms. For the fitting of the density, the A2 auxiliary function set was used [41]. The exchange-correlation potential was numerically integrated on an adaptive grid [42]. The grid accuracy was set to $10^{-5}$ in all calculations. The Coulomb energy was calculated by the variational fitting procedure proposed by Dunlap, Connolly and Sabin [43,44]. A quasi-Newton method in internal redundant coordinates with analytical energy gradients was used for the structure optimization [45]. The convergence was based on the Cartesian gradient and displacement vectors with a threshold of $10^{-4}$ and $10^{-3}$ a.u., respectively. These thresholds were decreased by a factor of 10 for the calculation of the vibrational spectrum, performed within the harmonic approximation. Finally, the barriers for trans to gauche conformational changes in butane and tetradecane were evaluated through transition state searches following the appropriate torsional mode (section IV.2.2).

## III- Models

We have explored the potential energy surface (PES) of the DMPC molecule. The starting structures (nearly 200) were generated from two different sources, namely (a) the geometries proposed in several earlier works [7,23-25], (b) the geometries obtained from systematic variations of all torsion angles of the molecule. We have only considered the conformations leading to a parallel stacking of the two alkyl chains, as found in the monomeric or micellar solutions [3]. The geometry optimizations of all degrees of freedom (348) of such generated structures, were seen to result in 50 minima which can be collected into several approximately isoenergetic groups. Among them, a group of fourteen was found to have the lowest energy, separated by about 3 kcal/mol from the others. These fourteen conformers lie within the range of 0.0 to 1.0 kcal/mol.

The DMPC molecule has been generally characterized as a head group or $\alpha$ chain that is connected through one $CH_2$ group (C1 carbon) to two glycerolipid $\beta$ and $\gamma$ chains. As clearly reviewed previously, these torsion angles have a major importance [1]. In this work, we follow the standard nomenclature defining the atoms and torsion angles, as shown in Figure 1. As an additional parameter, the dihedral angle $\phi$ = (O=C,C=O) has also been used (Figure 1).

In addition, for a better understanding of the torsional angles that control the orientations of different parts of the molecule, we have chosen to partition the molecule into four building blocks, labelled as head (H), neck (N), body (B) and tails. The head is built from the $N(CH_3)_3$-$CH_2$-$CH_2$-$PO_4$ atoms and is described by the $\alpha1$ to $\alpha6$ torsional angles. The neck includes the first $CH_2$ (standard nomenclature starts from this C1) as well as the next CH-$CH_2$ group (C2-C3) linking the two glycerolipid chains. The neck, characterized by the $\theta1$, $\theta2$ angles, defines the orientation of the head with respect to the glycerol groups and the tails. The body is formed by the next two O-CO-$CH_2$-$CH_2$ groups of the $\beta$ and $\gamma$ chains and is characterized by the $\theta3$, $\theta4$, $\beta1$ to $\beta4$ and $\gamma1$ to $\gamma4$ torsion angles. The tails consist of the two $\beta$ and $\gamma$ alkane chains and are defined by the $\beta_n$ and $\gamma_n$ torsions with $n\geq5$. For the search of the ground-state conformers, only trans $\beta$ and $\gamma$ angles ($n\geq5$ ) have been considered for the starting geometries, being known as the most stable rotational isomers for normal alkanes.

## IV- Results and Discussion

### IV.1 Analysis of the lowest-energy DMPC conformers

The fourteen lowest isoenergetic conformers can be subsequently grouped into two sets of structures which differ only in their β3 and γ3 values with respect to each other. This difference results in two mutual orientations of the carbon skeletons of the β and γ alkyl chains as shown in Figure 2. In the first group of seven lowest energy structures these two carbon skeletons are in two parallel planes (Fig 2a), whereas in the second half of the minimum energy conformers the two carbon skeletons are in two perpendicular planes (Fig 2b). For a given mutual orientation of the hydrocarbon chains ("parallel" or "perpendicular") the ground-state conformers, labelled in Table 1, arise from different couplings between head (H), neck (N) and body (B). The nomenclature adopted for these seven conformers is presented in Table1.

Among these seven lowest energy conformers (Table 2), we first notice two sets of θ3/θ4 values (body conformation) which are labelled as $B_A$ (θ3/θ4=180°/60°) and $B_B$ (θ3/θ4=60°/-60°). They correspond to the two experimentally proposed rotamers (A and B in ref. 3). At this point, it should be recalled that the stereochemistry around the C2-C3 bond leads to the relationship θ3-θ4=120°, whereas that around C1-C2 implies θ1-θ2=-120°. We notice that among the ground-state structures there are 6 $B_A$ and 1 $B_B$ conformations. Moreover, we observe that for each rotamer, β1-β4 and γ1-γ4 torsion angles can vary in a combined way yielding two different mutual orientations of the glycerol carbonyls, namely ϕ = 60° ($B_{A1}$/$B_{B1}$) and ϕ = 180° ($B_{A2}$/$B_{B2}$). The three body conformations $B_{A1}$, $B_{A2}$ and $B_{B1}$ are shown in Figure 3.

Two sets of α-chain torsions and three sets of θ1/θ2 angles, labelled as H1, H2 and N1, N2, N3, respectively, have been found. The torsion angles presented in Table 2 were averaged within a given set of these building blocks (H, N and B). Their maximum deviations are also reported in Table 2. The exact values of these dihedral angles, obtained for each conformer, are given in the supporting information.

*Body conformations*

In this section, we comment on the structural aspects of the DMPC body part. It is interesting to note that for all the conformers analyzed in this work (including higher energy structures discussed below) the γ2 and γ4 torsions are 180, and the β1 and β2 torsions are always nearly 120 and 180, respectively. The different conformers are produced by changes in γ1, γ3 and β3, β4 couplings. For example, in the case of $B_{A1}$ and $B_{A2}$ (conformers of the same rotamer but different orientations of C=O), the coupling between ($\gamma$1, $\gamma$3 = 90°, - 150°) and ($\beta$3, $\beta$4 = - 150°, 70°) gives $B_{A1}$ while, ($\gamma$1,$\gamma$3 = 110°, - 90°) and ($\beta$3, $\beta$4 = 180°, 175°) lead to the $B_{A2}$. In the case of $B_{B1}$, a change of the $\theta$3/$\theta$4 with respect to $B_{A1}$ and $B_{A2}$ necessitates a simultaneous changes in $\gamma$1, $\gamma$3, $\beta$3 and $\beta$4 in order to maintain the alkyl chains next to each other. Such coupled modulations of the diacylglycerol part were proposed in the earlier works analyzing the DMPC conformations in the liquid-crystalline phase [3].

In addition, we note that changes in the $\beta$3 and $\gamma$3 torsion angles (Table 2) lead to "parallel" and "perpendicular" mutual orientation of the two acyl chains (Figure 2). It is worth noting that the dispersion-type energies differ for these two chain arrangements. The optimum distance for "dispersion" between two normal alkane chains is about 4.2 Å, the value obtained for the "parallel" arrangement [37]. Indeed, the distance between the two alkane chains in the DMPC "parallel" conformers is also about 4.2 Å (Figure 2a) as compared to the distance of 4.8 Å (Figure 2b) in the DMPC "perpendicular" conformers. This correlates with a decrease of the "dispersion" stabilization by about 4.5 kcal/mol in the "perpendicular" conformation. However, in DMPC this effect is compensated by a structural rearrangement at the two O-C=O groups, which lowers their repulsion by increasing the C23…C32 distance from 4.2 to 5.5 Å, the next C24...C33 distance from 4.2 to 4.8 Å, and the other ones accordingly (Figure 2). The balance of the above two effects leads to similar total energies for the conformers with "parallel" and "perpendicular" tails. This conclusion is obtained computing the energies of the $H1N1B_{A1}$ with "parallel" and "perpendicular" tails, at their optimized structures, with and without the "dispersion" correction.

*Head and Neck conformations.*

Analysis of the head and neck conformations present in the lowest energy structures reveals that, in all cases, the head is bent towards either the $\beta$ or the $\gamma$ glycerol carbonyl. As proposed earlier [1,3-6] a bent head configuration should minimize the intramolecular electrostatic interactions in the polar region of the phospholipids. Indeed, all the lowest energy structures display the head and neck arrangements that minimize the internal electrostatic energy resulting from the repulsion between the phosphorous group and the carbonyl oxygen, on the one hand, and the attraction between the choline and carbonyl groups, on the other. These competitive interactions are accompanied by the modulation of the $\alpha$- chain torsion angles in order to preserve the internal hydrogen bond(s) between the methyl hydrogen(s) in choline and the nonester oxygen

(s) of the phosphate group (Figure 1). The P…N distance in the preferred conformers is 3.9 - 4.0 Å, agreeing well with those proposed by Hauser at al. [1]. It is worth noting that the higher energy conformers also have very similar P…N distances, showing that an optimal value for the P…N distance does not necessarily yield the most stable configuration. This choline-phosphate distance is determined by the α5 torsion angle value that is either -70° or 70° (Table 2). This result is in agreement with the experimentally reported values [1,7].

Given the constraint on α5, the α2, α3 and α4 angles combine with θ1/θ2, generating different ground-state structures by rotation around the C1 - C2 bond (Table 2, Neck conformation). The H1 conformation can combine with N1, N2 and N3 values, whereas H2 leads to ground-state isomers only when combining with N2 and N3. These head-neck couplings are also related to a particular body conformation. For example, the H1N1 coupled with $B_{A1}$, $B_{A2}$ and $B_{B1}$ bodies provides ground-state conformers, whereas the H1N1$B_{B2}$ is higher in energy. Similarly, only $H_1N_2B_{A2}$ is a ground-state conformer whereas the H1N2 combinations with $B_{A2}$, $B_{B1}$ and $B_{B2}$ lead to high energy conformations. One of the experimental DMPC structures proposed by Hauser et al.[1] has θ1, α2 and α3 values similar to those of the $H_1N_3$ conformer. However, in contrast to the commonly reported angle α1=180°, we find α1 = -100° in H1 and α1 = 105° in H2.

In order to bring out the factors behind the preferred combinations of the α, θ, β and γ torsions, we present in Table 3 characteristic geometrical parameters associated with the lowest and selected higher (by 3 kcal/mol) energy conformations.

The properties which display differences among the structures (Table 3) involve group distances and bond angles: the distances of the closest C=O oxygen to the N choline atom (N…C=O) and to the nonester phosphate O atom (P=O...O=C), respectively; the (O-P-O) and (C1-O-P) bond angles. These differences are due to the presence of different head conformations (H1 and H2) in the ground-state structures. For example, in the H1N1$B_{A2}$ conformer the (C=O....N) distance is 5.9 Å, whereas in the H2N3$B_{A1}$ structure the (C=O....N) distance is much shorter (3.9 Å). However, the (P=O...O=C) distance in H2N3$B_{A1}$ has also become shorter, as seen from Table 3. This closer position of the head with respect to the glycerol is moreover accompanied by a decrease of about 5° of the (O-P-O) and (C1-O-P) bond angles, yielding a larger strain in the H2 head. The difference between the H1 and H2 internal strains was evaluated by cutting out the heads of the optimized H1N1$B_{A2}$ and H2N3$B_{A1}$ geometries between the C1 and C2 atoms (Figure 1), and completing the C1 valence with a terminal hydrogen. This evaluation demonstrates that the H1 head is more stable than H2 by 5 kcal/mol, illustrating the H2 internal strain. The fact that H2 is closer to the glycerol moitie than H1 leads to a stronger choline-glycerol attraction, a stronger phosphate-glycerol repulsion and a larger internal head strain. Therefore, the total head-glycerol interaction in H2N3$B_{A1}$ can be estimated to be about 5 kcal/mol.

It is worth noting that all the ground-state conformers have similar Mulliken net charges (q) of the characteristic atoms: q(P) = 1.2, q(N) = -0.1, q(phosphate nonester oxygen) = -0.7, q (phosphate ester oxygen) = -0.5. The same is valid for the total net Mulliken charges of the choline and phosphate groups, evaluated at +0.6 and -1.2, respectively. The higher energy configurations, discussed below, have also very similar net atomic charges as those of the ground-state conformers.

## IV.2 Analysis of higher energy conformers

### IV.2.1 Conformers with all trans chains

Several sets of conformers lying in the range of 3 to 12 kcal/mol above the ground-state were also obtained from our calculations. Five structures, at 3 kcal/mol above the ground-state, are presented in Table 3 for comparison with the lowest-energy conformers. Among these five conformers we will first discuss three structures combining the H2 head and N1 neck with the $B_{A1}$, $B_{A2}$ and $B_{B1}$ bodies, i.e. H2N1$B_{A1}$, H2N1$B_{A2}$, H2N1$B_{B1}$. The same head and body combinations are found among the ground-state structures but associated with other necks than N1 (see Tables 1 and 2). The analysis of the geometrical parameters of these three higher energy conformers reveal that the (P=O…O=C) and (P…C=O) distances are shorter by 0.5-1 Å when compared with their values in the ground-state structures with the H2 head and N2 or N3 necks. On the other hand, the (N…O=C) distances as well as the (O-P-O) angles are very similar in both sets of the ground-state and higher energy conformers including the H2 head. As a consequence, the electronic repulsion between the P=O and C=O groups is increased and not compensated by an increased choline - carbonyl attraction or a decreased head strain. Keeping in mind that the H2N1 conformations are 3 kcal/mol higher in energy, one can therefore estimate the increased phosphate - carbonyl repulsion at about 3 kcal/mol.

Most of the DMPC molecular dynamics studies use the X-ray structures as starting geometries (DMPC1 and/or DMPC2) [1]. Therefore, we found it interesting to analyse their relative stability and properties at the QM level. The geometric properties reported in Table 3 were obtained after geometry optimization. We noted that the optimized torsional angles remain close to their starting values, in particular, α1 is close to 180° in DMPC1 and to 120° in DMPC2. The geometry optimization leads to an improved adjustment of the head folding over the β- chain glycerol, mainly through adjustments in the β1–β3 and γ1–γ3 torsion angles. However, the geometry optimization cannot decrease the choline – glycerol distances, which remain much larger than those found in the ground-state (by about 2 Å). An analysis of the physical origin of this result can be made by comparing the DMPC2 and $H1N1B_{A1}$ structures, since both have similar θ, β and γ torsional angles but opposite α1 to α5 angles (similar values but opposite signs). The existence of negative versus positive α torsional angles leads to symmetric heads with respect to a mirror plane. Changing α1 from 120° (DMPC2) to -100° ($H1N1B_{A1}$) generates another rotamer (rotation around the O11-C1 bond) with a different orientation of the head with respect to the glycerol backbone. Whereas the P-N direction is coplanar with the C21-C31 carbonyl carbons direction in DMPC2, it is about perpendicular in $H1N1B_{A1}$, implying a larger folding of the head in the latter, a shorter choline-carbonyl distance and a larger stability.

In fact, the optimum value of α1 depends on its combination with the neck torsional angles: H2 (α1 = 105°) combined with N1 (θ1 = 180°) and any of the bodies leads also, as described above, to a higher energy conformer, 3 kcal/mol above the ground-state.

### IV.2.2 Gauche conformations in the alkyl chains

It is well-known that n-alkanes exist in their ground-state as all-trans conformations. The transformation from trans to gauche requires the system to overcome a barrier corresponding to an eclipsed conformation of two C-C bonds. Using the methodology presented in section II, we find a barrier of 5.0 kcal/mol for butane. For hexane, the two barriers (rotating $CH_3$ and $CH_3$-$CH_2$) amount to 4.8 kcal/mol. A similar result (4.8 - 4.9 kcal/mol) is obtained for the tetradecane. Concerning the gauche isomers, their relative energy with respect to the all-trans conformer is also very similar from butane to hexane, with calculated values of 0.84 and 0.65 kcal/mol, respectively, for the first and second related gauche conformers. A value of about 0.75 kcal/mol has been measured from the butane gauche-trans interconversion in solid neon [46]. Interestingly, DMPC does not follow the behaviour of a $C_{14}$ n-alkane chain. Table 4 shows the evolution of the gauche-trans energy difference for DMPC conformers with one gauche transformation at various positions along the chain, compared with tetradecane. The torsions in Table 4 follow the nomenclature indicated in Figure 1 for the β and γ chains. The relative energies obtained for the β and γ chain values are averaged (they differ by less than 0.2 kcal/mol). Two conclusions can be drawn from these data: (i) the gauche conformers are less stable in DMPC than in the n-alkane of similar length; (ii) gauche transformations in the middle of the DMPC chains lead to less stable conformers. It is worth noting that the presence of two gauche transformations in the same chain or distributed one per chain lead to similar destabilizations of 1.8 – 2.0 kcal with respect to the all-trans structure.

IV.3 Calculated vibrational spectrum of DMPC and comparison with IR data

IR spectra provide information on the structure of the molecular functional groups and also on their interaction with the environment. In practice, the presence of different geometric characteristics for molecular conformers may be difficult to detect, due to large bandwidths and overlapping modes, which occur more often for large molecules with a large number of similar groups, for example the $CH_2$ groups in DMPC. The most recent studies on model and biological membranes were performed using Fourier transformed infrared (FT-IR) and isotopically labelled atoms, $^{2}H$ and $^{13}C$ more particularly, for given groups in the molecule. Deuterated phospholipids have been used to assign the C-H stretching bands in the phosphocholine group [47, 48] and also the frequency domains of the chain $CH_2$ wagging and rocking vibrations [9, 10]. Deuterated $CH_2$ vibrational frequencies were also analyzed to determine the presence of gauche chain conformers [11, 12]. Substitution of the $^{12}C$ carbonyl carbon of the β-chain by $^{13}C$ led to the unambiguous assignment of the carbonyl C=O and ester C-O stretching bands of the β and γ chains, revealing that the broad observed band was composed of two separate signals [9]. Despite the recognized red shift of about 20 $cm^{-1}$ for the antisymmetric O-P-O stretching ($\nu_a$) due to hydrogen bonding with water, some temperature dependence of ν(C=O) [9], most of the characteristic IR bands related with the lipid polar regions are similar in the gel and liquid-crystalline states [10]. In contrast, a significant blue shift is observed for C-H stretching frequencies when the temperature is increased above the phase transition temperature, which has been correlated with the appearance of gauche conformations in the acyl chains [9, 11, 12]. It is thus interesting to make a detailed comparison of the calculated vibrational spectrum of one of the DMPC conformers ($H1N1B_{A2}T1$, T = 0 K) with the experimental IR frequencies and assignments obtained for DMPC bilayers.

Prior to this analysis, it is worth noting that computed vibrational frequencies are very sensitive to the theoretical method used, in particular to the orbital basis set expansion and are generally shifted with respect to the experimental values. In order to estimate what difference can be expected for DMPC frequencies, the vibrational spectra of the gaseous butane and ethyl propanoate ($CH_3$-$CH_2$-CO-O-$CH_2$-$CH_3$) molecules have been calculated with the same methodology and compared with their experimental spectra [49]. This comparison shows that the calculated C-H stretching frequencies are overestimated by about 15 $cm^{-1}$ for the asymmetric stretchings of methyl and methylene groups, $\nu_a$($CH_3$ and $CH_2$), 8 $cm^{-1}$ for the symmetric stretching $\nu_s(CH_3)$ and 20 $cm^{-1}$ for $\nu_s(CH_2)$, the bending, wagging and rocking modes being in good agreement with experiment, within 5 $cm^{-1}$. For the propanoate, the calculated $\nu$C=O stretching at 1691 $cm^{-1}$ is strongly underestimated with respect to the1760 $cm^{-1}$ gaseous experimental value. The calculated $\nu$C-O ester is found 538 $cm^{-1}$ below $\nu$C=O, whereas the experimental difference is about 550 $cm^{-1}$. These errors, due to the method (orbital bases, functionals, zero temperature, harmonic approximation), are systematic and thus, once quantified, do not prevent a valid comparison with the experimental data.

The calculated vibrational frequencies and assignments are presented in Tables 5 and 6, whereas the calculated spectrum in the range 500 - 1800 $cm^{-1}$ and 2800 - 3100 $cm^{-1}$ is shown in Figure 3. The region below 500 $cm^{-1}$ contains very weak vibrational modes and the 1800 - 2800 $cm^{-1}$ region is free of signals. The reported experimental IR bands belong to the 900 – 1800 $cm^{-1}$ and 2800 – 3000 $cm^{-1}$ regions. Table 5 collects the calculated $\nu_a$ and $\nu_s$ C-H stretching frequencies, assigned to the $CH_3$ and $CH_2$ groups of the head, neck, body and chains of DMPC. FT-IR frequencies and assignments are also presented for comparison. The experimental assignments are based on the frequency shifts of different deuterated DMPC derivatives [48]. The highest calculated frequency corresponds to one of the asymmetric C-H stretching vibrations of the neck $CH_2$ groups. It is located in the small band visible at 3030 $cm^{-1}$ due to the choline methyl asymmetric C-H stretchings. The other neck $\nu_a$(C-H) frequencies at around 2965 are also hidden at the bottom of the very intense peak (maximum at 2940 $cm^{-1}$), which also includes the weak head $CH_2$ $\nu_a$(C-H) signal and the more intense $\nu_a$(C-H) modes of the end-chain $CH_3$ groups. Due to the overlap of all these vibrations, the neck $\nu_a$(C-H) bands could not be assigned experimentally. It is worth noting that the calculated $\nu_a$(C-H) bands of the chain $CH_2$ groups cover the 2910-2942 cm-1 range with three very intense vibrations at 2940, 2935 and 2925 cm-1. The comparison of these $\nu_a$(C-H) bands with experiment shows a very good agreement for the frequencies and similar assignments for the observed bands. The symmetric $\nu_s$(C-H) frequencies are found, as expected, in the low frequency side of the intense 3000 $cm^{-1}$ band. The strongest $\nu_s$ (C-H) band, in agreement with the experimental assignment is related with the chain $CH_2$ groups. Its 20 $cm^{-1}$ blue shift with respect to the observed peak is also in accordance with the butane and propanoic ester tests. Our calculations confirm the assignment of Gauger et al. [48] for the position of the hidden neck $CH_2$ $\nu_s$(C-H) mode and allow to assign the position of the choline $CH_3$ $\nu_s$(C-H) modes. It is worth noting that the 72 C-H stretching frequencies of the molecule are gathered, in both experimental and calculated spectra, between 2850 and 3030 $cm^{-1}$.

The other vibrations of the molecule are observed mainly between 700 and 1800 $cm^{-1}$. The calculated spectrum shows that all the frequencies below 500 $cm^{-1}$, which correspond to skeleton deformations and chain C-C-C-C torsions, have negligible intensities (<10). The main group vibrational modes are reported and assigned in Table 6, together with the available experimental data. Among them, the two chains $\nu$(C=O), the $PO_4$ vibrations and the rocking of the chain $CH_2$ are the most intense, in agreement with experiment. The carbonyl stretchings have been extensively studied experimentally for different phospholipids, in the gel and liquid-crystalline phases and with different hydration states [10]. The phase and hydration changes were shown to be accompanied with $\nu$(C=O) shifts, leading to the conclusion that the observed non-equivalence of the two C=O frequencies is associated with interactions with the environment, especially water. Our calculations show that the two carbonyls are structurally non equivalent and have thus two different $\nu$(C=O) bands, the $\gamma$ chain C=O at higher energy, with a difference of about 20 $cm^{-1}$, in good agreement with the experimental splitting of about 15 $cm^{-1}$. The $\beta$-chain carbonyl, with a 20 $cm^{-1}$ lower frequency, interacts substantially with the choline head group, as discussed above (IV.2.1). This interaction slightly increases the C=O bond length (~0.008 Å), implying the related frequency red shift. It is worth noting that the opposite effect is produced on the ester C-O bond length and frequency. Indeed, a weakening of the $\beta$ carbonyl bond is accompanied by a strengthening of the associated ester C-O bond. The difference between the $\beta$- and $\gamma$-chain C-O (ester) bond lengths (1.376 and 1.386 Å respectively) correlates thus with a higher frequency (1159 $cm^{-1}$) for the shorter β- chain C-O bond with respect to the γ- chain C-O (1124 $cm^{-1}$). The difference between calculated and experimental C=O and C-O bonds are similar with those reported above for the ethyl propionate test. This frequency region is rich in medium intensity bands in the experimental spectra, as well as in the calculated spectrum of Figure 3. We believe that this is the reason why only one ester C-O band has been assigned in the literature (Table 5).

The antisymmetric and symmetric stretchings, $\nu_a$(O-P-O) and $\nu_s$(O-P-O) of the phosphate group have been assigned to bands at 1255 and 1092 $cm^{-1}$ for lipids with $^{13}$C labelled carbonyls in the β- or γ- chains [13]. In hydrated non labelled phospholipid bilayers, the $\nu_a$(O-P-O) vibration is assigned to a band observed near 1220-1240 $cm^{-1}$, whereas the $\nu_s$(O-P-O) vibration is related to a band at about 1085 $cm^{-1}$ [10]. The calculated $\nu_a$(O-P-O) vibration at 1216 $cm^{-1}$ corresponds to a narrow and intense band. In contrast, the calculated $\nu_s$(O-P-O) at 1036 $cm^{-1}$ has a small intensity and mixed with head C-C stretchings. In fact, most of the stretching vibrations occurring between 1100 and 900 $cm^{-1}$ involve the combination of most head and body bonds, i.e. C-C, C-O and P-O, with the exception of the 973 $cm^{-1}$ asymmetric stretching of the choline group.

The other assigned bands of the experimental phospholipid IR spectra correspond to the bending, wagging and rocking $CH_2$ modes. In fact, the chain $CH_2$ bending and rocking modes are assigned to single bands, whereas the wagging vibrations are assigned to a series of weak regularly spaced bands [9, 10]. Our calculated spectrum reproduces very well these trends, with an intense band at 1445 $cm^{-1}$ for the chain $CH_2$ bending modes and small shoulders for head and neck $CH_2$ bendings. The wagging bands are indeed weak and spread among the region 1107-1220 $cm^{-1}$. The chain $CH_2$ rocking band at 720 $cm^{-1}$ is, in agreement with experiment, narrow and well visible. All the other bands between 1215 and 720 $cm^{-1}$ are less intense and they are mainly characteristic of stretching vibrations of the lipid backbone (see Table 5). One can however assign the asymmetric stretching modes of the C-$N^+(CH_3)_3$ choline group at 973, 939, 936 and 827 $cm^{-1}$ and their symmetric counterpart $\nu_s$ (-C-$N^+$-$(CH_3)_3$) at 660 $cm^{-1}$. These choline vibrations have not been assigned previously for phospholipids, except the 973 $cm^{-1}$band. Such vibrations have been observed for the more symmetrical tetramethyl ammonium (TMA) cation associated with different counter-anions with $\nu_a$ (N-$(CH_3)^{4+}$) at 947 $cm^{-1}$ for the TMA-Ce($SO_4$) crystal and at 946 and 959 $cm^{-1}$ for the TMA-$HF_2$ crystal and the $\nu_s$ counterparts at 764 and 759

cm$^{-1}$, respectively [51,52]. We see that the much less symmetrical choline group in DMPC has a larger $\nu_a$- $\nu_s$ splitting.

Finally, it is interesting to note the assignment in our caculated spectrum of the two intense bands at 673 and 617 cm$^{-1}$ to the asymmetric $\delta_a$ and symmetric $\delta_s$ deformation vibrations of the phosphate group, respectively.

The molecular calculated frequencies are thus in very good agreement with the experimental DMPC bilayer spectra, confirming mostly the assignments based on isotopically labelled molecules, allowing one to complete the assignments for hidden or low frequency vibrations. These results also show that the non-equivalence of the two glycerol vibrations originates from their structural difference, mainly associated with the choline – glycerol interaction. Interaction with water may increase the splitting but is not its fundamental origin.

Finally, we want to point out that the calculated spectrum of another conformer (H1N1B$_{A2}$T1), differing mainly in the body torsion angles, reproduces very similar signals, with a maximum shift of 10 cm$^{-1}$ for the related normal modes.

## V- Conclusions

In this study of the conformational properties of a DMPC molecule, we have shown that there is a substantial number of ground-state isomers, differing by selected torsion angles in the head, neck and/or body as well as by the relative orientation of the alkyl chains. We found 14 of them but there are most probably other possible combinations of $\alpha$, $\theta$, $\beta$ and $\gamma$ torsion angles leading to similar electronic energies. All these structures share a common geometric profile which ensures their stability: the attractive choline – glycerol group interaction is modulated by bond angle strain in the head and repulsive phosphate – glycerol effects. Finer energy tuning is found through choline – phosphate hydrogen bonding and damped dispersion attraction between the alkyl chains. The definition of this profile fits with most of the structural characteristics derived previously from the numerous NMR studies (spin – spin or dipolar couplings) of DMPC solutions or hydrated bilayers, showing that the existence of a single rigid conformer is not required for their interpretation. The assignments of the FT-IR spectra of various DMPC assemblies, achieved by using $^{2}$H and $^{13}$C substitutions, are confirmed by our calculated vibrational spectrum, which reproduces the characteristics of all typical vibrational modes, in particular the presence of two carbonyl stretching bands, related with the existence of a significant head – glycerol interaction. Conformational studies of other phospholipids are in progress.

Hence, our results support the conclusion that the intramolecular forces are preponderant in the determination of the phospholipid molecular structures. The fact that DMPC molecules retain their individual structure within assemblies has already been suggested in NMR studies [2,3]. Optimized dimeric structures, built from various DMPC conformers, show indeed that the substantial intermolecular electrostatic interactions between the monomers do not modify the individual monomer geometries (variations less than 3% for angles and torsions and less than 1% for bond lengths). The intermolecular effects certainly play a role in defining the mutual positions of the different conformers coupled inside the bilayer, or, eventually, in selecting some of them. Moreover, we can infer that DMPC and, eventually other phospholipid molecules, exist as various isomers with comparable energies, as is also underlined in lipid-binding proteins [50]. Finally, in a parallel study we have used Born-Oppenheimer Molecular Dynamics to show that the dynamics of the melting transition is also fundamentally a single-molecule process [51].

**Acknowledgements**: The Ambassade de France en Inde (New-Delhi) is gratefully acknowledged for a grant attributed to S.K. The bilateral project CNRS - Bulgarian Academy of Sciences is acknowledged for mission support to M.D.

Table1.The various possible combinations of the Head (H), Neck (N) and Body (B) present in the DMPC ground- state conformers.

| Conformatiion | Label |
|---|---|
| 1 | H1N1 $B_{A1}$ |
| 2 | H1N3 $B_{A1}$ |
| 3 | H2N3 $B_{A1}$ |
| 4 | H1N1 $B_{A2}$ |
| 5 | H1N2 $B_{A2}$ |
| 6 | H2N2 $B_{A2}$ |
| 7 | H1N1 $B_{B1}$ |

Table 2. Averaged torsion angles for the heads, necks and bodies of the DMPC ground-state conformers with “perpendicular“ alkyl chains. The maximum deviations between the averaged and exact torsion values are also given. The same combinations of heads, necks and bodies exist also for the conformers with “parallel” tails, their corresponding β3 and γ3 values are given in parenthesis.

| Label | Head Conformation | | | | | | | | | |
|---|---|---|---|---|---|---|---|---|---|---|
| | $\alpha_5$ | $\alpha_4$ | $\alpha_3$ | $\alpha_2$ | $\alpha_1$ | | | | | |
| **H1** | - 70±2 | 121±1 | 66±15 | 85±13 | - 100±15 | | | | | |
| **H2** | 70±1 | 49±2 | 45±4 | 178±7 | 105±3 | | | | | |
| | Neck Conformation | | | | | | | | | |
| | $\theta_1$ | $\theta_2$ | | | | | | | | |
| **N1** | 180±15 | - 75±10 | | | | | | | | |
| **N2** | - 60±15 | 60±15 | | | | | | | | |
| **N3** | 60±20 | 180±20 | | | | | | | | |
| | Body Conformation | | | | | | | | | |
| | $\theta_3/\theta_4$ | $\phi$ | $\gamma_1$ | $\gamma_2$ | $\gamma_3$ | $\gamma_4$ | $\beta_1$ | $\beta_2$ | $\beta_3$ | $\beta_4$ |
| **$B_{A1}$** | 180/60 | 60 | 90±5 | 175±5 | - 150±5 | 180±5 | 135±5 | 180±5 | - 150±5 | 70±5 |
| | | | | | (- 60±5) | | | | | |
| | | | | | | | | | (- 120±5) | |
| **$B_{A2}$** | 180/60 | 180 | 110±5 | 170±5 | - 90±5 | 180±5 | 120±10 | 175±5 | 180±5 | 175±5 |
| | | | | | (- 145±5) | | | | (145±5) | |
| **$B_{B1}$** | 60/- 60 | 60 | 110±5 | 185±5 | 170±5 | 180±10 | 120±10 | 175±5 | - 30±10 | 180±5 |

Table 3. Important intramolecular distances in Å and angles in degree of the conformers of the DMPC molecule obtained from the revPBE-LYP/DZVP geometry optimization.

| Conformer | (P=O)…(O=C) | N…O=C | O-P-O | P…C=O | H11…C=O<br>H12…C=O |
|---|---|---|---|---|---|
| Lowest Energy Conformers, ΔE=0.0 kcal/mol | | | | | |
| H1N1 $B_{A1}$ | 5.47 | 5.96 | 101 | 4.99 | 3.73<br>5.38 |
| H1N1 $B_{A2}$ | 5.52 | 5.86 | 101 | 4.95 | 3.26<br>5.41 |
| H1N2 $B_{A2}$ | 5.74 | 5.87 | 101 | 4.99 | 3.43<br>5.41 |
| H1N3 $B_{A1}$ | 5.51 | 5.75 | 100 | 5.48 | 3.42<br>5.56 |
| H1N1 $B_{B1}$ | 5.52 | 5.85 | 101 | 4.97 | 3.21<br>5.42 |
| H2N2 $B_{A2}$ | 4.04 | 4.04 | 95 | 5.51 | 4.87<br>2.76 |
| H2N3 $B_{A1}$ | 4.50 | 3.89 | 95.5 | 5.46 | 5.10<br>3.60 |
| Higher Energy Conformes, ΔE=3.0 kcal/mol | | | | | |
| H2N1 $B_{A1}$ | 3.38 | 4.12 | 94.4 | 3.78 | 4.71<br>2.55 |
| H2N1$B_{A2}$ | 3.44 | 4.14 | 95 | 3.93 | 5.15<br>3.10 |
| H2N1$B_{B1}$ | 3.49 | 4.17 | 94.8 | 3.89 | 5.17<br>3.11 |
| *DMPC1 | 6.17 | 8.14 | 100 | 6.05 | 5.93<br>7.80 |
| *DMPC2 | 6.12 | 7.10 | 100 | 5.37 | 3.59<br>5.75 |

* DMPC1 and DMPC2 are the structures, optimized at revPBE-LYP/DZVP starting from the torsion angles given in ref. 7 for the single crystal DMPC conformers.

Table 4: Relative energies[a] of gauche conformers (with respect to the all-trans) for gauche transformations at various positions in the alkyl chains[b] for DMPC and tetradecane ( n-$C_{14}H_{30}$)

| βn /γn trans to gauche | DMPC | Tetradecane |
|---|---|---|
| n=14 | 0.94 | 0.65 |
| n=12 | 1.26 | 0.46 |
| n=10 | 2.07 | 0.48 |
| n=8 | 0.72 | 0.48 |
| n=6 | 0.70 | 0.46 |

[a] kcal/mol

[b] Each gauche conformer is obtained changing one βn or one γn angle from the trans to gauche (60°) value, followed by a complete geometry optimization. The change in energies with respect to trans are similar for both chains and the averaged value is indicated for DMPC. The nomenclature for βn and γn is given in the Figure 1

Table 5. Calculated C-H stretching frequencies of DMPC (H1N1B$_{A2}$T1), assignments to the head, neck and chains vibrational modes[a] and comparison with experimental results[d]

| Vibrational Modes[b] of groups | Calculated Frequencies ($cm^{-1}$) (calc. Intensity)[c] | Experimental Frequencies[d] | Experimental Assignments[d] |
|---|---|---|---|
| νa(C-H) neck $CH_2$ | 3032 (15) | | |
| νa(C-H) head choline $CH_3$ | 3030-2976 (10-40) max. 3030 | 3026.5 | νa(C-H) choline $CH_3$ |
| νa(C-H) neck $CH_2$ | 2972 (10), 2967 (83), 2959 (30) | | n.a. |
| νa(C-H) head $CH_2$ | 2957 (20), 2950 (6) | 2955.5 | νa(C-H) head $CH_2$ |
| νa(C-H) end chains $CH_3$ | 2944 (122) | 2955 | νa(C-H) end chains $CH_3$ |
| νa(C-H) chains CH2 | 2942-2910 (4-652) 2940 (360), 2935 (119), 2925 (652) | 2919 | νa(C-H) chains CH2 |
| νs(C-H) head $CH_2$ | 2919 (56) | | |
| νs(C-H) head + neck $CH_2$ | 2914 (21), 2913 (26), 2907 (2), 2906(24) | 2886 | νs(C-H) head CH2 |
| νs(C-H) head $CH_3$ | 2912(15), 2909 (22) | | n.a. |
| νs(C-H) chains $CH_2$ | 2898 (337), 2897(87) 2888 (110), 2876 (75) | 2849.5 | νs(C-H) chains $CH_2$ |
| νs(C-H) end chains CH3 | 2887 (32) (in the 2898 peak) | 2872 | νs(C-H) end chains $CH_3$ |
| νs(C-H) head $CH_3$ | 2860 (143) | | n.a. |

[a]head CH2 groups include C11 and C12; neck CH2 groups include C1, C2 and C3; body CH2 groups include C22 and C32; chain CH2 include all C2x and C3x with x>2; end chain CH3 include the two terminal chain methyl groups.
[b]a : asymetric, s : symmetric group vibration.
[c]Calculated intensities are given in km/mol; band maxima values visible on the Figure ? are underlined.
[d] values and assignments from ref. 48; underlined values correspond to the band maxima presented in ref. 44; n.a. are not observed or not analyzed in ref. 48.

Table 6 . Calculated vibrational frequencies of DMPC (H1N1B$_{A2}$T1) and assignments to the head, neck, body and chains vibrational modes[a] for the 1800-500 $cm^{-1}$ region[b]; Experimental values for various DMPC systems are given for comparison.

| Vibrational Modes [b,c] of groups | Calculated Frequencies ($cm^{-1}$) (calc. Intensity)[d] | Experimental Frequencies |
|---|---|---|
| ν(C=O) γ chain | 1696 (203) | 1740[e] |
| ν(C=O) β chain | 1676 (149) | 1725-1727[f]<br>1693[e] ($^{13}$C=O lipid) |
| | | |
| δa $CH_3$+ bend. $CH_2$ head | 1492-1447 (15-28) | |
| bend. chains $CH_2$ | 1445 (80), 1438 (43), 1436 (16), 1432 (35) 1425 (19),1418(22) max. 1445 | 1467[e]<br>1462-1474[f] |
| δa $CH_3$ head | 1437 (36) | |
| δs head $CH_3$ | 1394-1384 (<10) | 1375[f] |
| wagg. head + neck $CH_2$ | 1371-1359 (<10) | |
| δs end chains $CH_3$ | 1355 (10), 1348 (<5) | 1375[f] |
| wagg. head + neck + body $CH_2$ | 1351-1348 (<5) | |
| wagg. chains $CH_2$ | 1345-1200 (<5) | 1300-1200[e]<br>1345-1195[f] |
| wagg. head $CH_3$ + twist. head $CH_2$ | 1290-1265 (9-10) | |
| wagg. neck + head + chains $CH_2$ | 1282-1220 (<5) | |
| νa O-$PO_2$-O head | 1216 (80) | 1220-1240[f]<br>1255[e] ($^{13}$C =O lipid) |
| wagg. chains CH2 | 1220-1107 (1-10) | |
| ν(O-C) body β chain | 1159 (28) | 1177[e] |
| ν(O-C) body γ chain | 1124 (27) | 1158[e] ($^{13}$C =O lipid) |
| ν(C-C) neck | 1110-1104 (8-9) max. 1110 | |
| ν(C-C) head + neck + body + chains (some bands include C-O-P stretch.) | 1065 (19), 1064 (27), 1035 (11), 996 (13), 980 (12), 964 (19) max. 1065 | |
| νs O-$PO_2$-O head involve also the O-C and C-C | 1036(11), 1032 (<5) | 1092[e] ($^{13}$C =O lipid)<br>1085[f] |
| ν(O=P-O-C-C) head | 1006 (32) | |
| ν(O-C) + C-C head | 997 (37), 995 (13)<br>max 997 | |
| ν(C-C) chains | 996 (13), 980 (12),<br>964 (19) | |
| νa (-C-$N^+$-(CH3)3) head | 973 (11) | 970[e] |
| νa(-C-$N^+$-(CH3)3) head | 939 (9), 936 (18) | |
| ν(C-C) head | 936 (18), 923 (7),<br>917 (7)<br>max. 939 | |
| ν(C-C) end chains | 864 (5), 862 (8) | |
| νa (-C-$N^+$-(CH3)3) head | 827 (16) | |
| rock. $CH_2$ head + chains | 728 (14), 718 (54)<br>max 720 | 720-730[f] |
| δ O-C=O | 697-689 (3-10) | |
| δa (O-PO2-O) head | 673 (138) | |
| νs (-C-$N^+$-(CH3)3) head | 660 (<5) | |
| δs (O-PO2-O) | 617 (52) | |
| δ (OCC)+ δ (CCC) + δ (COP) | 547 (23), 535 (6)<br>max 547 | |

[a]head CH2 groups include C11 and C12; neck CH2 groups include C1, C2 and C3; body CH2 groups include C22 and C32; body OC=O groups include C21, C31, O21,O31; chain CH2 include all C2x and C3x with x>2; end chain CH3 include the two terminal chain methyl groups.
[b]a : asymetric, s : symmetric group vibration.
[c]ν(X-Y): stretching of bond X-Y ; bend.: bending; wagg.: wagging; twist.: twisting; rock. rocking; δs and δa symmetric and asymmetric deformation modes. The δs mode for CH3 is also called the umbrella mode.
[d]Calculated intensities are given in km/mol; band maxima values visible on the Figure ? are underlined.
[e] ref. 13.
[f] ref. 9 and 10.

Figure Captions

Figure 1: Nomenclature of the atoms and torsion angles of the DMPC molecule.

Figure 2: Comparison of the conformers : (a) “parallel” tails; (b) “perpendicular” tails.

Figure 3: Illustration of three conformers with H1 head and N1 neck but different bodies (a) $B_{A1}$; (b) $B_{A2}$; (c) $B_{B1}$.

Figure 4: Calculated vibrational spectrum of the DMPC molecule. The signals are simulated using lorentzian functions with half-width value of 6 cm-1.

Figure 1

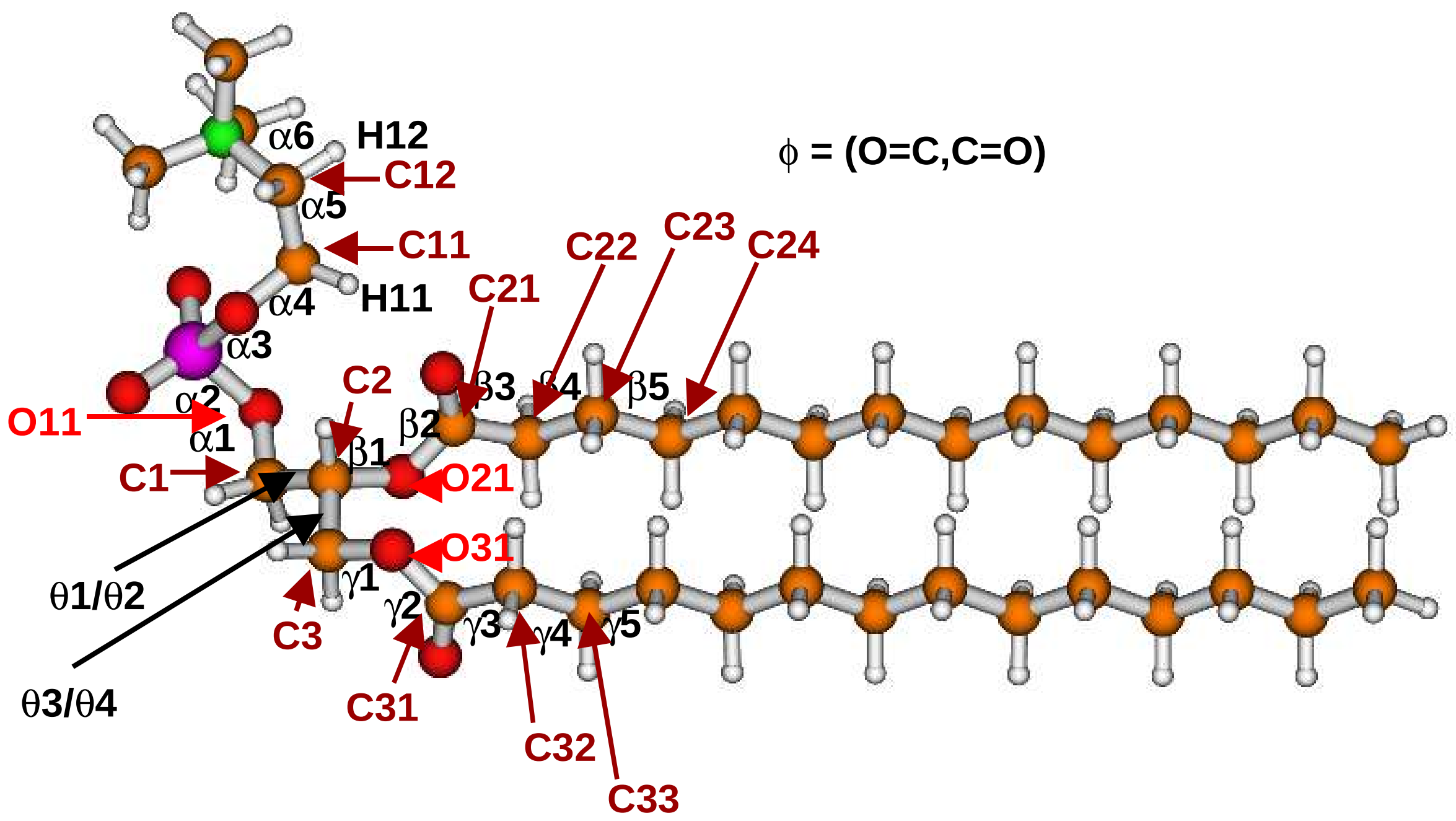

Figure 2

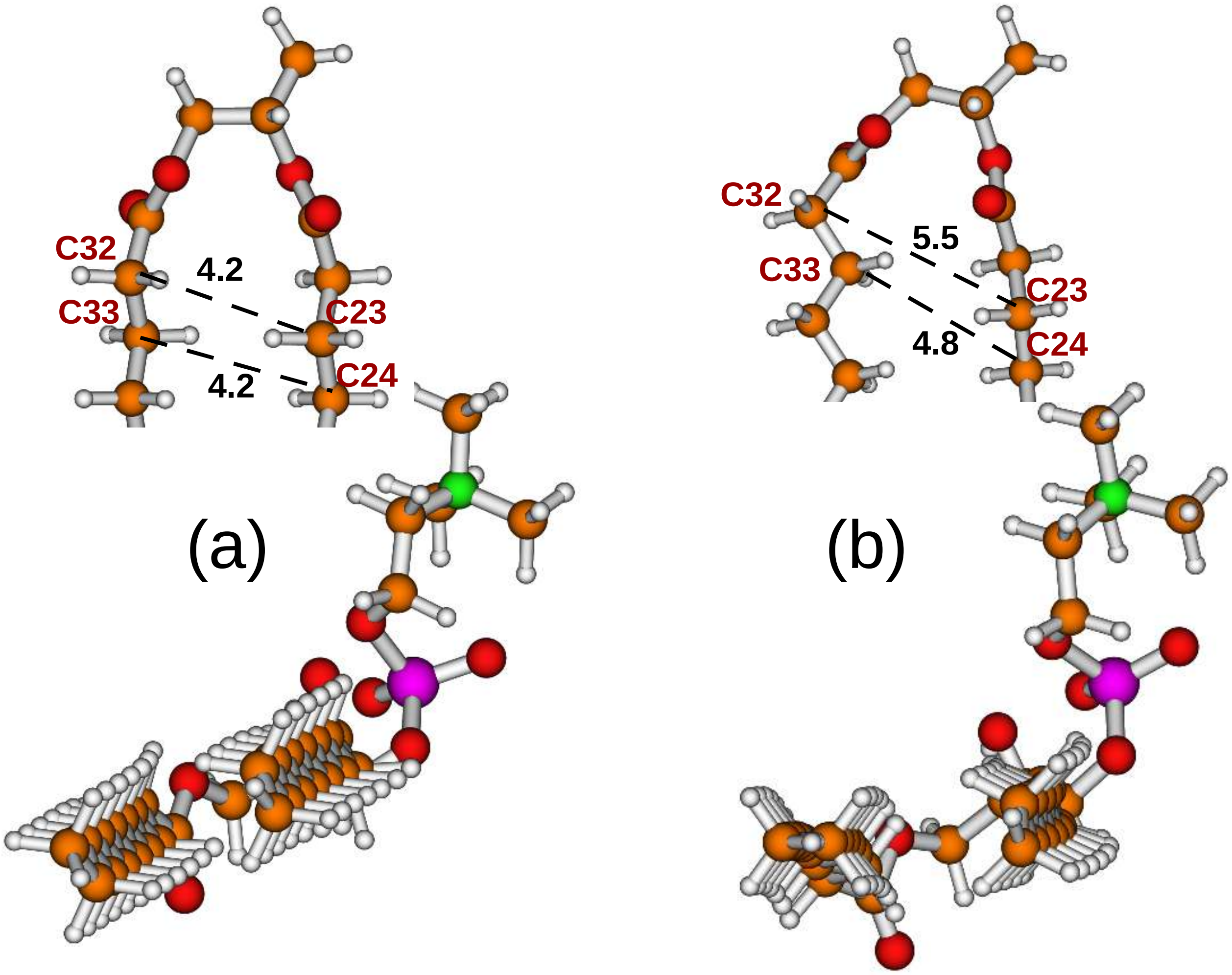

Figure 3

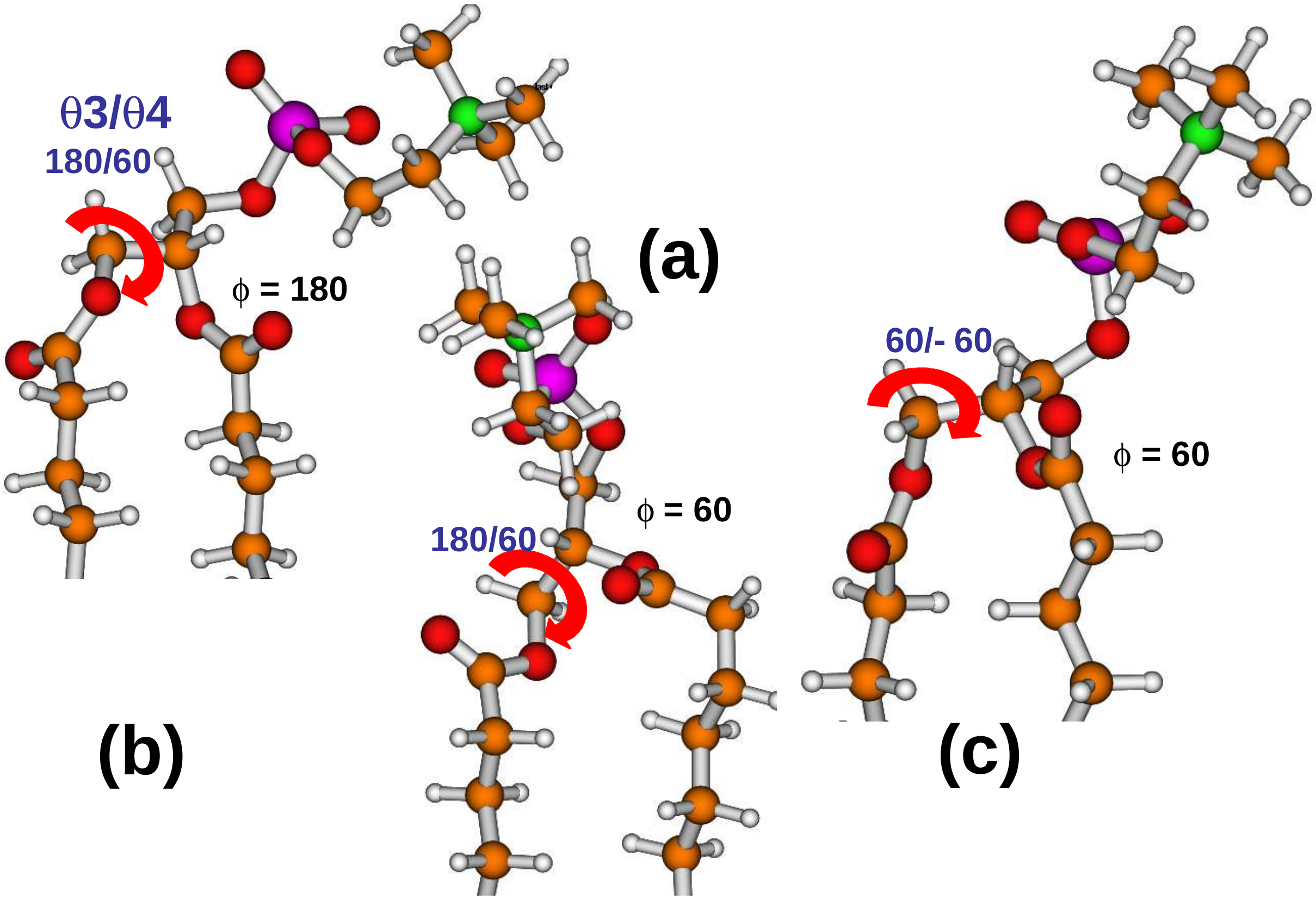

Figure 4

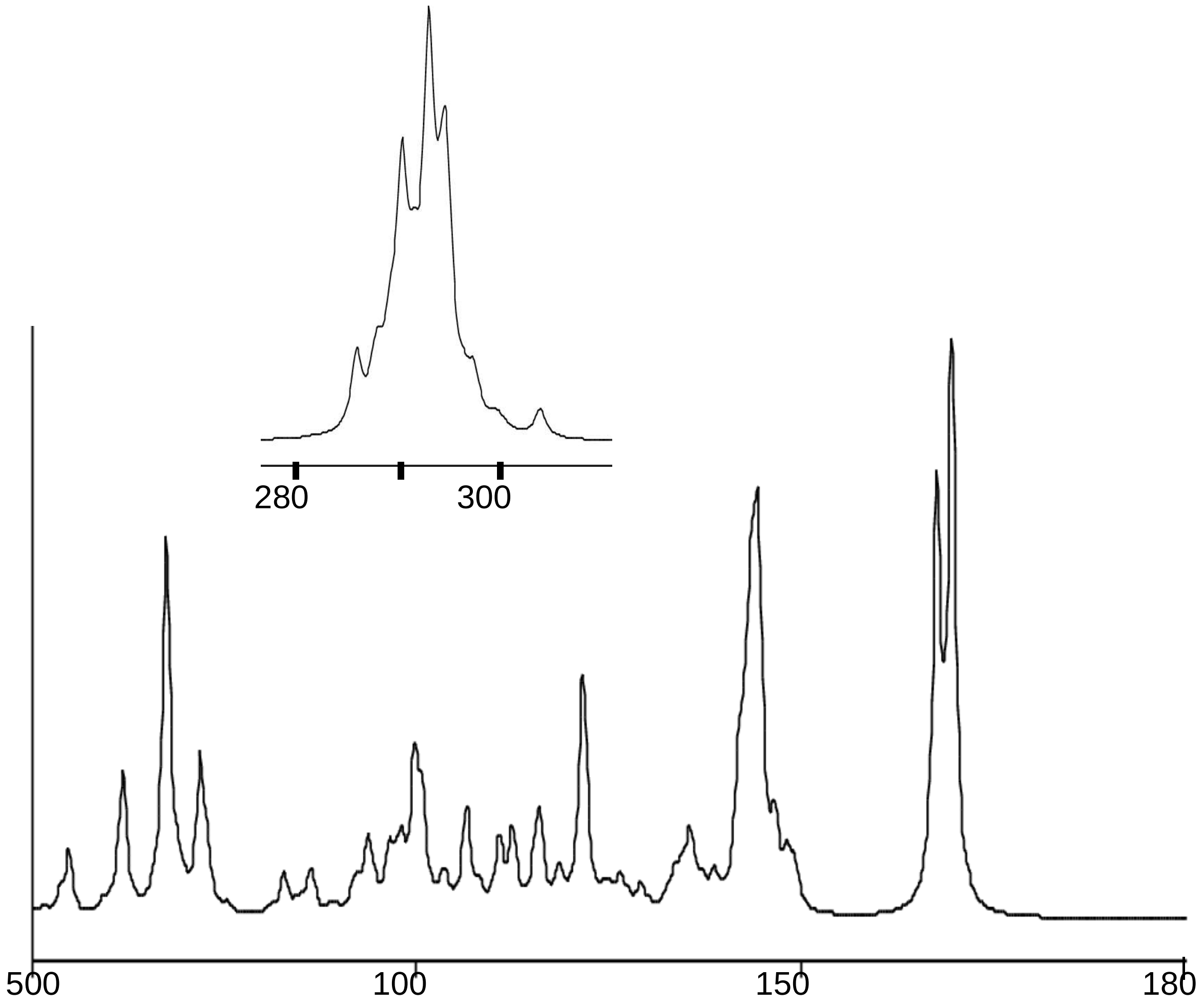